\documentclass{aa}
\usepackage{graphicx}
\usepackage{txfonts}
\begin{document}
\title{
Spectroscopic signatures of magnetospheric accretion in Herbig~Ae/Be stars. \\
I.~The case of HD\,101412.\thanks{Based on observations made with ESO Telescopes at the La Silla Paranal Observatory
under programme IDs~087.C-0124(A), 088.C-0218(A,B,C,E), 090.C-0331(A), and 092.C-0126(A).
} }

\author{M.~Sch\"{o}ller\inst{1}
   \and
M.~A.~Pogodin\inst{2}
    \and
J.~A.~Cahuasqu\'{i}\inst{3}
    \and
N.~A.~Drake\inst{4,5}
    \and
S.~Hubrig\inst{6}
    \and
M.~G.~Petr-Gotzens\inst{1}
    \and
I.~S.~Savanov\inst{7}
    \and
B.~Wolff\inst{1}
    \and
J.~F.~Gonz\'{a}lez\inst{8}
    \and
S.~Mysore\inst{1}
    \and
I.~Ilyin\inst{6}
    \and
S.~P.~J\"arvinen\inst{6}
    \and
B.~Stelzer\inst{9}
}

\institute{
European Southern Observatory,
Karl-Schwarzschild-Str.~2,
85748~Garching,
Germany\\
\email{mschoell@eso.org}
    \and
Central Astronomical Observatory at Pulkovo,
Pulkovskoye chaussee 65,
196140~Saint Petersburg,
Russia
    \and
I.~Physikalisches Institut,
Universit\"at zu K\"oln,
Z\"ulpicher Str.~77,
50937~K\"oln,
Germany
    \and
Saint Petersburg State University,
Universitetskiy pr.~28,
198504~Saint Petersburg,
Russia
    \and
Observat\'orio Nacional/MCTI,
Rua General Jos\'e Cristino~77,
CEP 20921-400,
Rio de Janeiro, RJ,
Brazil
    \and
Leibniz-Institut f\"ur Astrophysik Potsdam (AIP),
An der Sternwarte~16,
14482~Potsdam,
Germany
    \and
Institute of Astronomy, Russian Academy of Sciences, Pyatnitskaya~48, 119017~Moscow, Russia
    \and
Instituto de Ciencias Astron\'omicas, de la Tierra y del Espacio (ICATE),
5400~San~Juan,
Argentina
    \and
INAF - Osservatorio Astronomico di Palermo,
Piazza del Parlamento~1,
90134~Palermo, Italy
}

\authorrunning{Sch\"{o}ller et al.}
\titlerunning{Magnetospheric accretion in Herbig Ae/Be stars}

   \date{Received ????? ???, ???; accepted ????? ??, ???}

% \abstract{}{}{}{}{}
% 5 {} token are mandatory

  \abstract
{
Models of magnetically-driven accretion and outflows reproduce many observational properties of T\,Tauri stars.
This concept is not well established for the more massive Herbig Ae/Be stars.
}
{
We intend to examine the magnetospheric accretion in Herbig Ae/Be stars and
search for rotational modulation using spectroscopic signatures,
in this first paper concentrating on the well-studied Herbig~Ae star HD\,101412.
}
{
We used near-infrared spectroscopic observations of the magnetic Herbig Ae star HD\,101412 to
test the magnetospheric character of its accretion disk/star interaction.
We reduced and analyzed 30 spectra of HD\,101412, acquired with the CRIRES and
X-shooter spectrographs installed at the VLT (ESO, Chile).
The spectroscopic analysis was based on the \ion{He}{i}~$\lambda$10,830 and
Pa$\gamma$ lines, formed in the accretion region.
}
 {
We found that the temporal behavior of these diagnostic lines in the near-infrared spectra of HD\,101412 can be
explained by rotational modulation of line profiles generated by accreting gas with a period $P = 20\fd53\pm 1\fd68$.
The discovery of this period, about
half of the magnetic rotation period  $P_{\rm m} = 42\fd076$
previously determined from measurements of the mean longitudinal magnetic field,
indicates that the accreted matter falls onto the star in
regions close to the magnetic poles intersecting the line-of-sight two times during the rotation cycle.
We intend to apply this method to a larger sample of Herbig Ae/Be stars.
}
{}


   \maketitle

\section{Introduction}
\label{sect:intro}

Herbig Ae/Be stars (HAeBes) are pre-main-sequence (PMS) objects  with pronounced
emission line features and an infrared (IR) excess indicative of dust in their
circumstellar (CS) disks (Herbig \cite{Herbig1960};
Finkenzeller \& Mundt \cite{FinkenzellerMundt1984};
Th\'e et al.\ \cite{The1994}).
It is now recognized that these stars are intermediate-mass analogues of T\,Tauri stars,
but with convectively stable interiors that do not support dynamo action as found in the
fully convective T\,Tauri stars (Gullbring et al.\ \cite{Gullbring1998}).
For this reason, unlike for the T\,Tauri stars,
strong magnetic fields of the order of 1\,kG are usually not expected in HAeBes.
On the other hand, in recent years a number of magnetic studies revealed that some Herbig Ae/Be
stars have globally organized magnetic fields of the order of 100\,G
(Hubrig et al.\ \cite{Hubrig2004};
Wade et al.\ \cite{Wade2005};
Alecian et al.\ \cite{Alecian2008};
Hubrig et al.\ \cite{Hubrig2009};
Alecian et al.\ \cite{Alecian2013};
Hubrig et al.\ \cite{Hubrig2013};
Hubrig et al.\ \cite{Hubrig2015}).
From detailed magnetohydrodynamical models, it is expected that magnetic fields in
low-mass PMS objects  funnel material from the disk towards the star and launch a
collimated bipolar outflow (Shu et al.\ \cite{Shu2000}).
The star/CS interaction in classical T\,Tauri stars is well described
by the magnetospheric accretion (MA) model (Bouvier et al.\ \cite{Bouvier2007}),
where the field truncates the disk at a distance of between five and ten stellar radii.
However, it is still unclear how well this model can be applied
to HAeBes, whose magnetic fields are roughly one order of magnitude weaker.

Recently, Cauley \& Johns-Krull (\cite{CauleyJohnsKrull2014})
presented results of an analysis of the \ion{He}{i}
$\lambda$10,830 profile morphologies for a significant sample of more than 50 HAeBes
in a wide range of spectral types.
They concluded that objects of early B types show no sign of the MA process.
The matter infall from their disks onto the star takes place near the equatorial plane.
On the other hand, the \ion{He}{i}
$\lambda$10,830 profile shape in the spectra of objects of late B and A types
indicates that they are surrounded by magnetospheres, but with radii much
smaller than in the case of T\,Tauri stars.
This result is not surprising, taking
into account the lower values of their magnetic fields (Hubrig et al.\ \cite{Hubrig2015}).
Clearly, important diagnostics of the star/CS interaction region of individual objects are accessible by
examining the temporal behavior of near-IR spectral lines of these objects.

In our studies, we aim to investigate the accretion process and test the
applicability of the MA model to selected Herbig Ae/Be stars with previously detected
magnetic fields, concentrating in this first paper on the well-studied Herbig~Ae star HD\,101412.
Our method is based on monitoring the variability
detected in the red part of line profiles originating in (or close to) the
region of the star/CS interaction.
If the orientation of the disk deviates from an edge-on orientation, then the detected variability can be considered as
a signature of the accretion flows intersecting the line-of-sight at intermediate and high latitudes.
This can take place only for MA accretion, when
the accreted material is carried out from the equatorial plane along closed magnetic
field lines inside the magnetosphere to higher latitudes.

We also intend to search for rotational modulation in spectral line
profiles. If the star has a significant magnetosphere and the magnetic axis is not
aligned with the rotation axis, the accreted flow will be governed by the
magnetic field inside the magnetosphere and the accretion shock on the stellar
surface near the magnetic pole will be observed as an azimuthal inhomogeneity.
Such an inhomogeneity rotates together with the star and modulates the line shape with a
period equal to the rotation period of the star.

More specifically, we investigate the variability observed in two near-IR
lines, \ion{He}{i}~$\lambda$10,830 and Pa$\gamma$ (at 10,938\,\AA{}), in the spectrum of
the strongly magnetic Herbig Ae star HD\,101412.
The important role of these  lines in probing
the structure of the accretion region of PMS objects has already been discussed by
Edwards et al.\ (\cite{Edwards2006}).

Hubrig et al.\ (\cite{Hubrig2010}) studied spectra of HD\,101412 obtained with UVES and HARPS
and identified resolved magnetically split lines indicative of a variable magnetic field
modulus, changing from 2.5 to 3.5\,kG.
Such a strong field is typical for T\,Tauri stars, but is rarely measured in HAeBe stars.
A study of the magnetic variability of HD\,101412 found a cyclical variation of
the mean longitudinal magnetic field $\left<B_{\rm z}\right>$ with an amplitude of
$A_{\left<B_{\rm z}\right>} = 465\pm27$\,G around a mean value of $\overline{\left<B_{\rm z}\right>} = 9 \pm 18$\,G
(Hubrig et al.\ \cite{Hubrig2011}).
HD\,101412 rotates very slowly, with a projected rotation velocity of $v\sin i\approx 3$~km\,s$^{-1}$
(Cowley et al.\ \cite{Cowley2010}).
Using the stellar fundamental parameters and the detected magnetic rotation period
of $P_{\rm m} = 42\fd076 \pm 0\fd017$,
the inclination angle of the rotation axis relative to the line-of-sight was estimated as
$i= 80 \pm 7\degr$ , and
the angle between the magnetic and the rotation axis was determined as $\beta = 84
\pm 13\degr$ (Hubrig et al.\ \cite{Hubrig2011}).
This means that the magnetic axis lies close to the  plane of the equatorial disk,
and that the regions close to the magnetic poles, where the accreted matter falls onto the star,
intersect the line-of-sight two times during one rotation period.

\section{Observations and data reduction}
\label{sect:obs}

The spectra  of HD\,101412 were acquired with 
CRIRES (CRyogenic high-resolution InfraRed Echelle Spectrograph)
in short wavelength ranges around the \ion{He}{i}~$\lambda$10,830 and Pa$\gamma$ lines
with a spectral resolution of $R\sim100,000$,
and using X-shooter ($R\sim 11,000$) 
to obtain spectral data simultaneously over the entire spectral range from the
near-UV to the near-IR in three different arms.
Both instruments are operated by the European Southern Observatory (ESO)
on the Very Large Telescope (VLT) on Cerro Paranal, Chile.
In total, 30 spectra were obtained from 2011 to 2014.
The full list of observations is presented in Table~\ref{tab:tableA1},
where Col.~1 gives the modified Julian date (MJD) at the middle time of observation, Col.~2 the instrument used,
Col.~3 the signal-to-noise ratio (S/N) reached in the continuum
between the \ion{He}{i}~$\lambda$10,830 and Pa$\gamma$ lines,
and Col.~4 the phase corresponding to the ephemerides presented by Hubrig et al.\ (\cite{Hubrig2011}).
A S/N of 100--500 was achieved for all spectra observed.
Both CRIRES and X-shooter spectra were reduced using the respective ESO pipeline.
The normalization of the spectra near Pa$\gamma$ was
carried out using a photospheric synthetic spectrum calculated with the code
SYNTH3 (Kochukhov \cite{Kochukhov2007}).
We assumed the atmospheric parameters $T_{\rm eff}$ = 8300\,K and $\log g =3.8$
(Cowley et al.\ \cite{Cowley2010})
and made use of the VALD atomic line database (Kupka et al.\ \cite{Kupka1999}).

%--------------------------------------------------------------------
\section{Spectroscopic signatures of magnetospheric accretion}   %Sec 3
\label{sect:analysis}

\begin{figure} %Fig 1
\centering
\includegraphics[width=82mm]{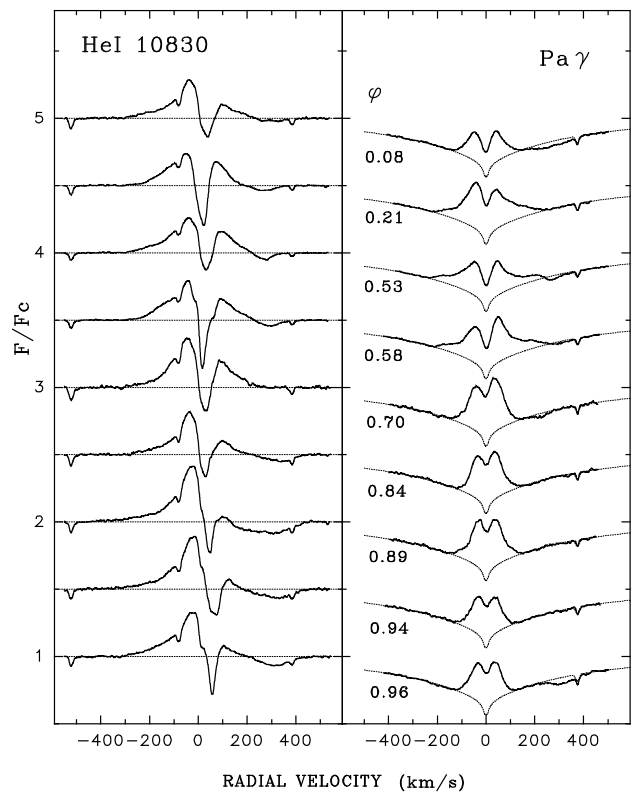}
\caption{
Line profiles of the  \ion{He}{i}~$\lambda$10,830 and Pa$\gamma$ lines in
HD\,101412 CRIRES spectra.
Rotation phases of the observations assuming the magnetic rotation
period $P_{\rm m}=42\fd076$ are presented close to the Pa$\gamma$ lines.
From top  to bottom, these phases correspond to observations obtained at
MJD\,55662.082, 55667.174, 55681.023, 55683.058, 56361.178,
56367.283, 55654.085, 55656.062, and 56372.185 (see Table~\ref{tab:tableA1}).
The epoch MJD\,52797.4 corresponds to phase
$\varphi=0$, as presented in Fig.~4 of Hubrig at al.\ (\cite{Hubrig2011}).
Synthetic Pa$\gamma$ profiles assuming the atmospheric parameters $T_{\rm eff}$ = 8300\,K and
$\log g =3.8$ are indicated by the dotted lines.
}
\label{fig:f1}
\end{figure}

The profiles of the \ion{He}{i}~$\lambda$10,830 and Pa$\gamma$ lines observed in the
CRIRES spectra of HD\,101412 are presented in Fig.~\ref{fig:f1};
a presentation of the X-shooter data can be found in Fig.~\ref{fig:xshooter_set}.
The \ion{He}{i} line appears as an emission profile with two separate redshifted absorptions.
The first of them, narrow and deep, shows a small velocity shift towards the red, typically at locations
between 20 and 70\,km\,s$^{-1}$.
The second absorption in the red is much wider.
It is rather flat and demonstrates a large velocity shift,
starting as low as 100\,km\,s$^{-1}$ and ending as high as 500\,km\,s$^{-1}$.
Similar wide redshifted absorptions are observed also in the Pa$\gamma$ profiles
on a few observing epochs.
All components in both line profiles are variable in intensity, shape and velocity.
Only the central absorption component in the Pa$\gamma$ profile shows no velocity shift.

The behavior of the near-IR lines in the spectrum of HD\,101412 is closely related
to the edge-on orientation of the accretion disk and the matter flows falling
onto the star from the inner edge of the disk.
Applying the magnetic field model of Hubrig et al.\ (\cite{Hubrig2011}), it follows
that the accretion onto the star occurs as a flow with two components related to
the two magnetic polar regions.
The broad redshifted absorption is formed in the
infalling flows screening the stellar disk near these magnetic polar regions.
The central absorption originates from the inner part of the disk.
The high-temperature region of the \ion{He}{i} formation is 
geometrically thinner than in the case of Pa$\gamma$.
It covers the volume where the accretion is just beginning, whereas the central Pa$\gamma$ absorption is generated in
a spatially more extended region, where the accreted flows are not yet significant.
An illustration of the principal components in a typical MA model is given, for
example, in Camenzind (\cite{Camenzind1990}).

It can be seen in Fig.~\ref{fig:f1} that the width of the emission
profiles of the \ion{He}{i} line are systematically larger than those of the
Pa$\gamma$ line.
This phenomenon can be a result of additional emission that is
present in emission wings of lines, connected with the stellar wind and the accretion flows
at higher latitudes, which do not screen the stellar disk due to the edge-on
orientation of the object.
According to the MA model, for a wind being driven along
open magnetic field lines, the field enforces corotation out to approximately
the Alfv\'{e}n radius $R_{\rm A}$ (Cauley \& Johns-Krull \cite{CauleyJohnsKrull2014}).
Inside $R_{\rm A}$, the
outflowing gas is accelerated by the magnetic centrifuge and becomes less dense due
to the open field configuration and mass conservation.
A high-velocity
wind  of low density at a distance near $R_{\rm A}$ is much better registered in the
resonance \ion{He}{i} line than in the subordinate Pa$\gamma$ line, which is very
sensitive to the gas density.
Emission in Pa$\gamma$ can appear only in gas with a density sufficiently
higher than that in the remote wind.

Assuming the presence of two streams passing the line-of-sight and screening the star two times
during one rotation period, we expect two times an increase of the broad redshifted
absorption in the line profiles and a decrease of the intensity of the emission components.
This takes place because screening the star by a stream leads to absorption of stellar
radiation by the infalling matter.
The brightness of the stream itself in the considered
lines is lower than that of the stellar disk.
Thus, the absorbing effect
from a stream passing is sufficiently stronger than the effect from its additional
emission.
As a result, the broad absorption components have to appear at large positive
velocities, where there is no emission from the disk.
Further, a decrease in the intensity of the emission profile is observed
at velocities where the disk emission overlaps with the radiation of star and stream.

\begin{figure}  %Figure 2
\centering
\includegraphics[width=67mm]{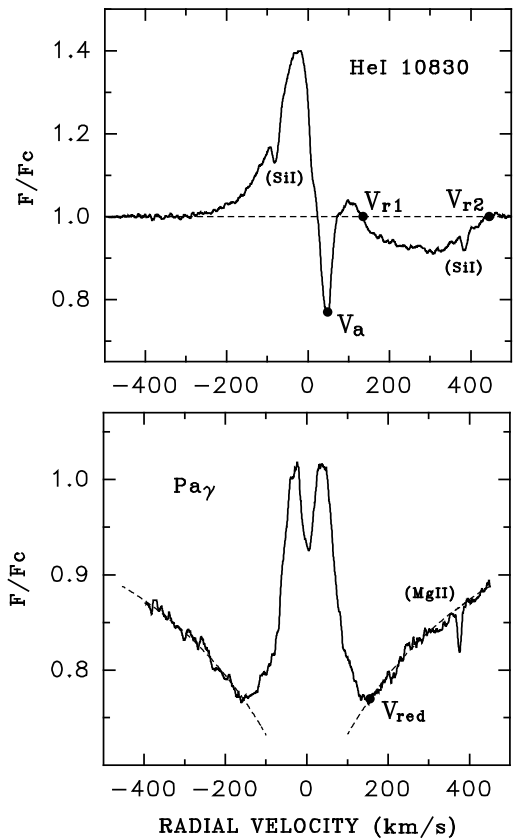}% mnr-2.eps
\caption{Spectral parameters of the \ion{He}{i}~$\lambda$10,830 and Pa$\gamma$
line profiles used in the quantitative analysis.}
\label{fig:f2}
\end{figure}

To investigate the temporal behavior of the spectral lines, we used specific line
parameters characterizing the change in the broad red absorption component and
the intensity variability of the line emission.
$v_{\rm r1}$ and $v_{\rm r2}$ are the
velocities of the blue and the red edges of the broad redshifted absorption of the
\ion{He}{i} line at the continuum level.
In the case of Pa$\gamma$, this absorption
component is not present in all observations.
For this line, we used the parameter $v_{\rm red}$,
which is the velocity of the red edge of the emission profile.
We assume that
this parameter is an analog of $v_{\rm r1}$ when the broad red absorption is not
detectable in our spectra and manifests itself as a depression of the red emission wing.
The parameter EW is the equivalent width of an emission profile above the atmospheric background
(determined as $F_{\rm line}/F_{\rm cont}-1$).
Additionally, we used the parameter $v_{\rm a}$, which is the velocity of the deep
central absorption of the \ion{He}{i} line, which is formed at the high-temperature
inner boundary of the disk and is an indicator of the origin of the accretion process.
The exact locations of these parameters are illustrated in Fig.~\ref{fig:f2}.
The measured values of all line parameters, together with the measurement accuracies
obtained at different dates, are presented in Cols.~5--10 of Table~\ref{tab:tableA1}.
The determination of the EW errors follows Smith et al.\ (\cite{Smith1995}).
The errors in velocity are deduced from a wavelength calibration term and
twice the error of the gradient of the line profile.
With CRIRES, we obtain errors of 20\,m\AA{} for the EW and between 1 and 30\,km\,s$^{-1}$
for the velocities.
For X-shooter, errors are typically a factor of 2 to 3 larger.

\begin{figure}  %Figure 3
\centering
\includegraphics[width=67mm]{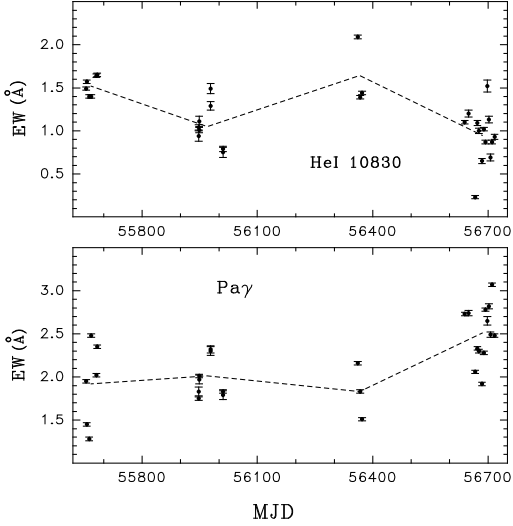}
\caption{
Temporal behavior of the EW of the \ion{He}{i}~$\lambda$10,830 and Pa$\gamma$ lines.
Linear trends between the time intervals of the observations are indicated by the dashed lines,
connecting the mean values observed in each epoch.
Error bars are indicated for all values.
}
\label{fig:f3}
\end{figure}

The assumption of the presence of rotating streams as a cause for the observed
variability of the line parameters implies the existence of correlations between
their variations.
One can expect a direct correlation between the EWs of the
\ion{He}{i} and Pa$\gamma$  lines, and between the EWs of both lines and
the parameters $v_{\rm r1}$(\ion{He}{i}) and $v_{\rm red}$(Pa$\gamma$).
An inverse correlation is expected between all these parameters and  $v_{\rm r2}$(\ion{He}{i}).
Before studying such correlations, we must eliminate long-term variabilities that
are not connected with the rotation of the stream.
The temporal behavior of the EW for both IR lines is illustrated in Fig.~\ref{fig:f3}.
A systematic difference is detected between values
of EW(\ion{He}{i}) obtained in the four observing periods MJD\,55650\,--\,55690,
MJD\,55940\,--\,56010, MJD\,56360\,--\,56370, and MJD\,56630\,--\,56720.
In the case of the EW(Pa$\gamma$), a similar
difference was observed only for the values observed in the fourth observing period.
No long-term variations were detected in the temporal behavior of the other line parameters. 
To eliminate the effect of long-term variability of the EWs,
which was assumed to be a linear trend between the four observing periods,
all values corresponding to the first three periods were divided by the ratio of the mean values
of the EWs calculated separately for the first three periods and that for the fourth
period.
As a result, four relevant correlations have been revealed
and the corresponding correlation coefficient $r$ determined:
\begin{eqnarray}
v_{\rm r1}(\ion{He}{i}) \, {\rm vs.} \, EW(\ion{He}{i}):         & r = +0.65\pm0.10, \\
v_{\rm r1}(\ion{He}{i}) \, {\rm vs.} \, v_{\rm r2}(\ion{He}{i}): & r = -0.73\pm0.08, \\
v_{\rm r1}(\ion{He}{i}) \, {\rm vs.} \, v_{\rm red}(Pa\gamma):   & r = +0.64\pm0.11, \\
v_{\rm red}(Pa\gamma)   \, {\rm vs.} \, EW(Pa\gamma):            & r = +0.70\pm0.09.
\end{eqnarray}
with the error of $r$ determined by $(1-r^2)/\sqrt{N}$ and $N$ the number of the measured values
(see equation 26.24 in Kendall \& Stuart \cite{KendallStuart1961}).

The existence of such correlations indicates the presence
of accreted matter between the star and the observer.
The discovery of such correlations would be enough to confirm the MA character of the accretion in an
object with an intermediate disk orientation.
But in the case of an edge-on oriented object like HD\,101412, where the accretion
process takes place practically in the equatorial plane, such correlations can occur
not only as a result of the presence of rotating streams, but also due to a change of parameters of
the accretion process in absence of a magnetosphere.
Only the detection of a rotational modulation of the line profiles can be considered a convincing
signature of the magnetospheric accretion.

%-------------------------------------------------------------------
\section{Periodicity search using the spectroscopic signatures}  %sect 4
\label{sect:period}

As outlined above, it is important to search
for signatures of rotational modulation in the different line profile
parameters to confirm the existence of a cyclic variability with a period near 
half the magnetic rotation period of HD\,101412 detected by Hubrig et al.\ (\cite{Hubrig2011}),
that is with a period of about 21\,d.

From a simple geometric consideration, variability related to rotational
modulation is expected to show a sine character if $\beta + i < 90\degr$,
when only one magnetic pole is visible to the observer during the rotation period.
For other cases, when both poles are observed over the rotation cycle, the
character of the modulation can be more complex and it is possible to describe it by a
function with several sinusoidal harmonics.
We note that we do not intend to
investigate this character in detail since our data set is not large enough for
such an analysis.
We try only to confirm a periodicity of the parameter variations and
to estimate the value of the period.
For this reason, we used the simplest method of
harmonic analysis with only one harmonic.
As we show below, this method allows us to
achieve more precise results than other methods not connected with a harmonic
analysis, for example, the method of Laefler \& Kinman (\cite{LaeflerKinman1965}).

\begin{figure}  %Figure 4
\centering
\includegraphics[width=0.24\textwidth]{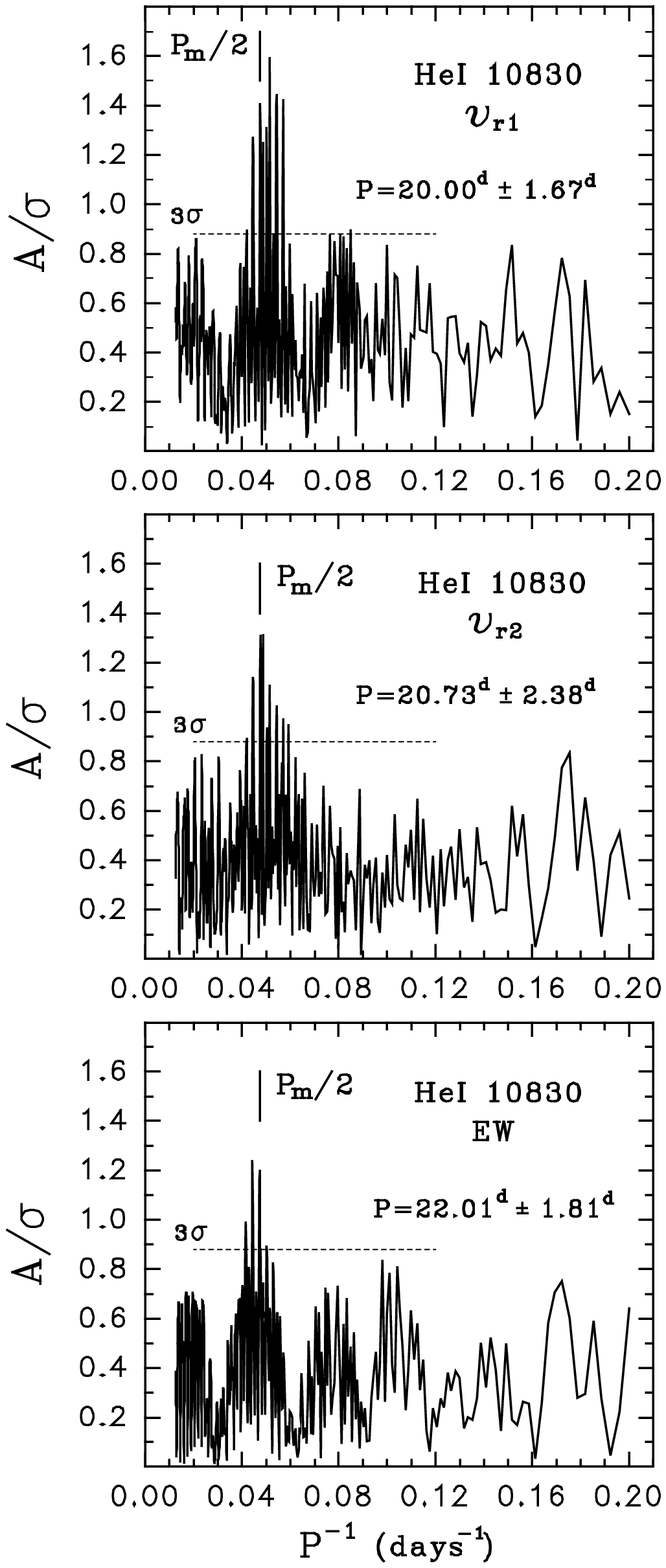}
\includegraphics[width=0.24\textwidth]{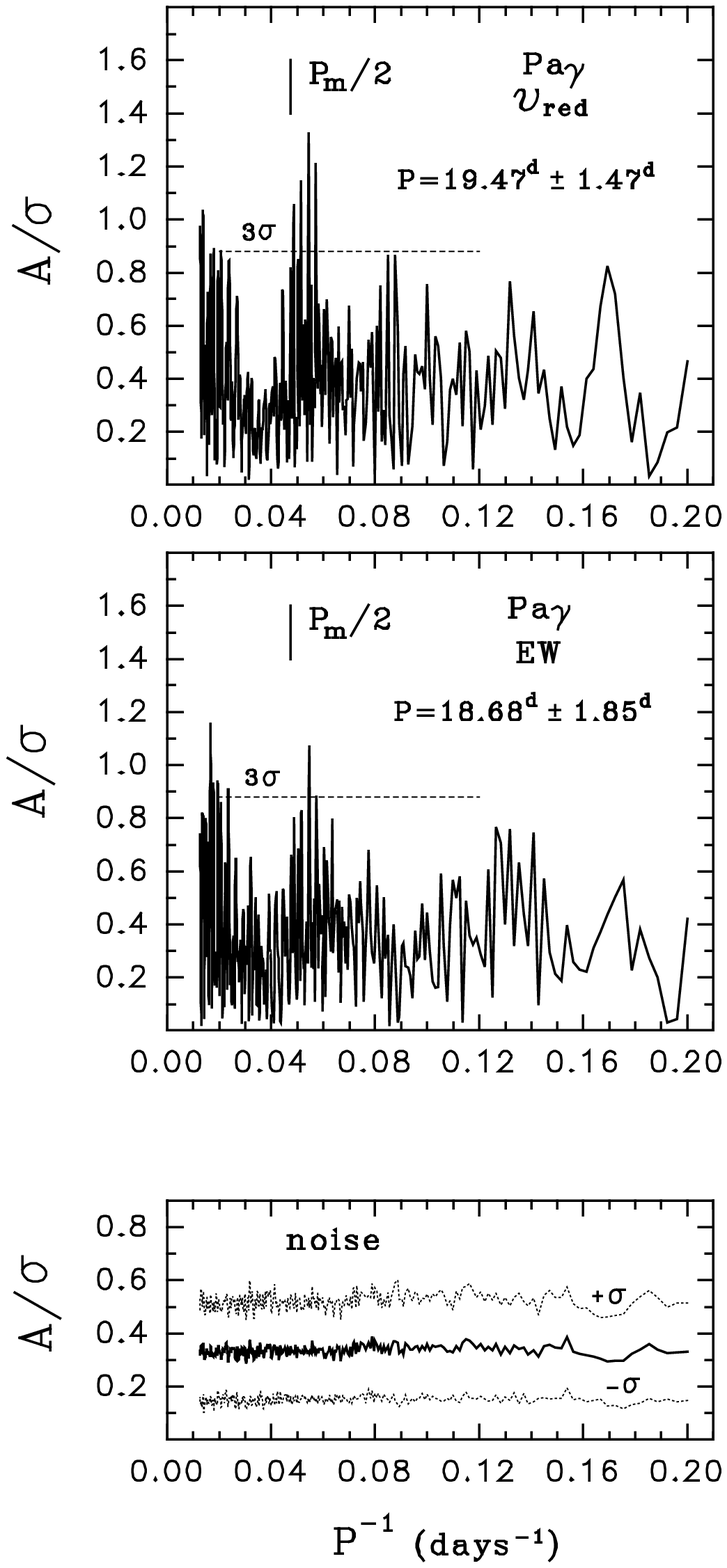}
\caption{
Different line parameters' $A/\sigma$ periodograms as defined in Sect.~\ref{sect:analysis}.
Significance levels of 3$\sigma$ are indicated by the dashed lines.
Short vertical lines indicate the value corresponding to half of the magnetic rotation period ($P_{\rm m}/2 = 21\fd038$).
The detected period values and their errors are given in each plot.
}
\label{fig:f4}
\end{figure}

Our search for periodicity is based on fitting phase dependencies for each value of the trial
period $P$ with a sine for a range of 5 to 80\,days, in steps of 0.1\,days.
The parameters of the sine, amplitude, constant coefficient, and initial phase,
were determined using the ordinary least-square method for each value of $P$.
We applied this method to the diagnostic line parameters introduced in the previous section.
The periodogram for $v_{\rm a}$ turned out not to be informative.
This parameter traces the region of the inner disk where the accretion starts,
which might be viewed differently for the two magnetic poles due to the disk not being fully edge-on
and the magnetic field not exactly perpendicular to the rotation axis.
The periodograms constructed for the other five parameters are shown in Fig.~\ref{fig:f4}.
$A/\sigma$ is the ratio between the amplitude of the sinusoid and the standard deviation
of the residuals of the sine function fit for a given $P$.
To determine the window function and to
estimate the significance level of the separate peaks in the periodograms, we also calculate 
the noise periodogram,
following closely the methods employed by Zechmeister \& K\"urster (\cite{ZechmeisterKuerster2009})
and Alvarado-G\'omez et al.\ (\cite{AlvaradoGomez2015}).
It was constructed by substituting the
line parameter values in the temporal sequence by a set of random numbers and the
calculation of a periodogram with the same series of dates.
More than 200 reiterations of this procedure
allowed us to generate a mean noise periodogram
and the standard deviation of an individual periodogram relative to the mean
(Fig.~\ref{fig:f4}, plot on bottom right).

A peak near half the magnetic rotation period $P_{\rm m}$/2 is seen in
all periodograms at a significance level higher than 3$\sigma$.
Each peak is rather wide and consists of several narrow local peaks.
Such a structure is the result of
a) the rather low number of observations
and b) the non-uniform distribution of all values over the observing dates,
with the temporal sequence divided into four separate groups (see Fig.~\ref{fig:f3}).
The positions of the local peaks inside each wide peak allow us to estimate the mean
value $P$ and the standard deviation for the given peak, which are indicated in
Fig.~\ref{fig:f4}.

To test our method of period determination, we carried out independent period
estimations using the standard Lomb normalized periodogram (LNP) based on an IDL (Interactive Data Language)
routine following Press et al.\ (\cite{Press1992}).
This analysis led to practically
identical periodograms including positions and relative amplitudes of all local peaks
inside the main wide peaks.

The most significant periods are the periods determined using the parameters of
the \ion{He}{i} line (Fig.~\ref{fig:f4}, left panels).
In the periodograms calculated for
Pa$\gamma$ (Fig.~\ref{fig:f4}, upper two panels on the right), a period near $P_{\rm m}$/2 is not so
obvious, some other periods at significant levels are also seen at $P\sim50^{\rm d}$
and $P\sim8^{\rm d}$.
Since the window function is rather smooth for our temporal sequence
(Fig.~\ref{fig:f4}, bottom right), these
additional periods cannot be a result of the specific data distribution over time.
Sometimes such period-artifacts can appear accidentally, as a result of poor
statistics.
We tested whether several periods present in the periodograms constructed
for the Pa$\gamma$ parameters are independent from each other.
The periodogram calculation was repeated, but a sine corresponding to $P\sim21^{\rm d}$ was
first subtracted from the temporal sequence.
It turned out that the periods at $P\sim50^{\rm d}$ and $P\sim8^{\rm d}$ disappeared too.
This means that the periodograms contain artifacts related to poor statistics.
In this case, the repetition of the period appearance in all periodograms can be a criterion of its validity.
The only real period is that near $P_{\rm m}/2$.

We also obtained the mean periodogram, averaging the periodograms from all five parameters.
As a result, we found the period of variation of the spectroscopic
data $P = 20\fd53\pm 1\fd68$, which is in good agreement with half the magnetic rotation
period $P_{\rm m}/2 = 21\fd038$.
For comparison, a similar estimate made with the Laefler-Kinman method
resulted in $P = 20\fd49\pm 2\fd48$.

\begin{figure}  %Figure 5
\centering
\includegraphics[width=0.24\textwidth]{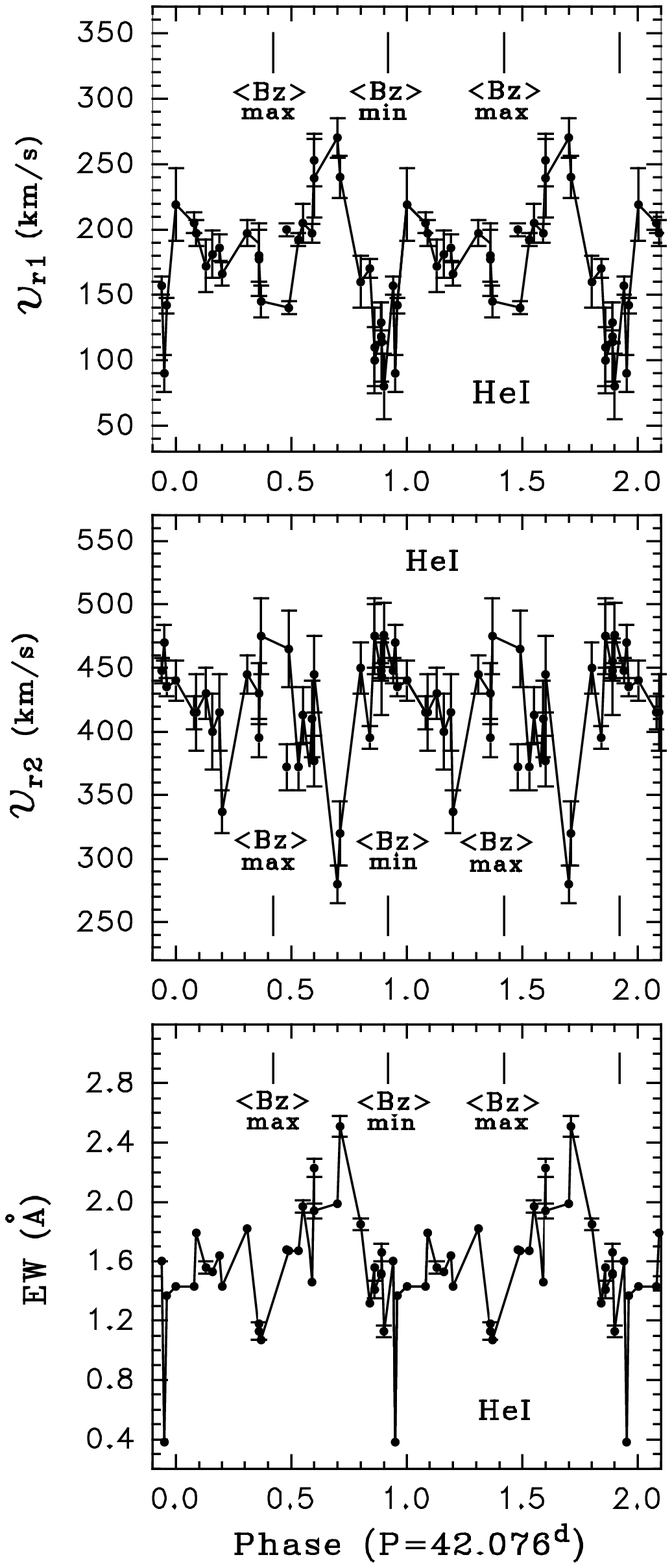}
\includegraphics[width=0.24\textwidth]{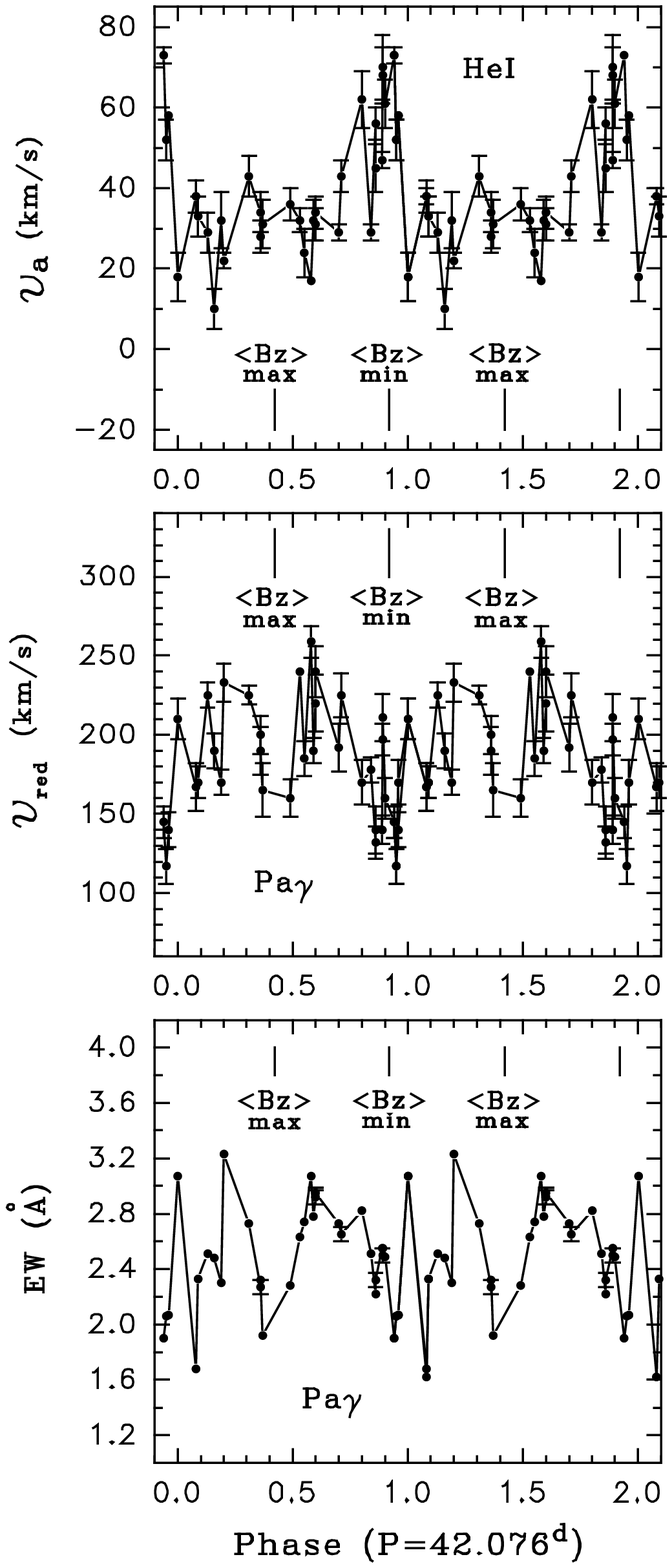}
\caption{
Phase dependencies of different line parameters over the magnetic rotation
period $P_{\rm m}= 42\fd076$.
The initial phase $\varphi = 0$ corresponds to MJD\,52797.4.
Error bars are given for each value.
}
\label{fig:f5}
\end{figure}

Fig.~\ref{fig:f5} illustrates the phase dependencies of the different line parameters
constructed for $P_{\rm m} = 42\fd076$ and the initial phase $\varphi = 0$ at
MJD\,52797.4 taken from Hubrig at al.\ (\cite{Hubrig2011}).
Phases where $\left<B_{\rm z}\right>$ reaches its maximum and minimum values are
marked in the figure.
One can see that the value of the $v_{\rm r2}$(\ion{He}{i})
parameter reaches its maximum just at the phases where $\left<B_{\rm z}\right>$
is at its minimum or maximum.
$v_{\rm a}$(\ion{He}{i}) shows a maximum only during the phase when $\left<B_{\rm z}\right>$
has a minimum, and with a smaller amplitude than $v_{\rm r2}$(\ion{He}{i}).
As the MA model predicts,
the values of such parameters as $v_{\rm r1}$(\ion{He}{i}), $v_{\rm
red}$(Pa$\gamma$), and EW (for both lines) become minimal at phases where
$\left<B_{\rm z}\right>$ reaches its minimum or maximum values.

We conclude that the phase dependencies of all chosen line parameters are in good agreement
with an MA model and with the magnetic field topology and orientation suggested by Hubrig
et al.\ (\cite{Hubrig2011}).
The phases when $\left<B_{\rm z}\right>$ takes its maximum and minimum values
correspond to the times when the magnetic poles are close to the line-of-sight.
At these phases, the matter flows move exactly away from the observer,
screening the star.
This results in a decrease of the EW values of the spectral lines
and a growth of the broad redshifted absorptions followed by an increase of
$v_{\rm r2}$(\ion{He}{i}) and $v_{\rm a}$(\ion{He}{i}) and a decrease of
$v_{\rm r1}$(\ion{He}{i}) and $v_{\rm red}$(Pa$\gamma$).
Interestingly, during our study of
the rotational modulation, we noticed that the value scatter in a number of phase
dependencies is much larger than the observational errors indicated in Table~\ref{tab:tableA1},
especially for the EW.
This means that the line parameters exhibit
additional variability not directly connected to the rotational modulation.
It is likely to be the result of a change in the accretion regime on different timescales.

%------------------------------------------------------------------
\section{Conclusions}
\label{sect:conclusions}

\begin{figure}
\centering
\includegraphics[width=0.45\textwidth]{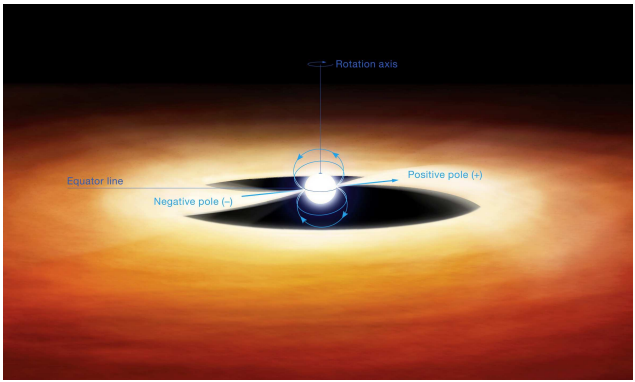}
\caption{
Artist's impression of the magnetospheric accretion in HD\,101412,
looking at the magnetic equator (corresponding to rotation phase 0.25).
The viewing and obliquity angles follow the model determined
by Hubrig et al.\ (\cite{Hubrig2011}).
The exact size and shape of both the disk and the accretion
streams are for illustration only, since we lack constraints
on these parameters.
}
\label{fig:sketch}
\end{figure}

According to Romanova et al.\ (\cite{Romanova2003,Romanova2004})
and Romanova \& Owocki (\cite{RomanovaOwocki2015}), magnetospheric accretion is a complex process,
and the interaction between the inner disk matter and the stellar magnetosphere
depends on a number of factors, such as the star's rotation period,
the structure of the stellar magnetic field, the size of the magnetosphere,
the diffusivity at the disk-magnetosphere boundary, properties of the accretion disk, and other elements.
Numerical global 3D MHD simulations of accretion onto stars with different
tilt angles of the dipole field have been performed by Romanova et al.\
(\cite{Romanova2003,Romanova2004}).
The authors showed that for dipole inclination angles larger than 60$^\circ$, matter accretes
in two streams that follow paths to the closest magnetic pole.
The streams have a shape different from those at small inclination angles and come to the star
near the equatorial plane.
The accretion rate is smaller for aligned dipoles than for tilted dipoles.
Further, the variability of different spectral lines is expected to depend on the density, temperature,
and velocity distributions along the line of sight to the star.
Since the \ion{He}{i} $\lambda$10,830 line probes inflow (accretion) and outflow (winds)
in the star-disk interaction region of accreting T\,Tauri and Herbig Ae stars
(Edwards et al.\ \cite{Edwards2006}; Fischer et al.\ \cite{Fischer2008}),
it can be successfully used to study the influence of magnetic field topologies on the star-disk interaction.
Revealing the relations between the
mass accretion rate and the magnetic field geometry is very promising since
they can constrain the predictions of theoretical studies of magnetospheric
accretion and wind launching models.

The results of our spectroscopic study of the strongly magnetic Herbig Ae star
HD\,101412 show that the temporal behavior of its near-IR lines
\ion{He}{i}~$\lambda$10,830 and Pa$\gamma$, originating from the region of the star/CS interaction,
can be successfully explained in the framework of a magnetospheric accretion model with the
geometry of the magnetic field suggested by Hubrig et al.\ (\cite{Hubrig2011}).
In Fig.~\ref{fig:sketch}, we present an artist's impression of the topology of the MA in HD\,101412,
as seen by an observer when looking onto the magnetic equator. 
The average period $P = 20\fd53\pm 1\fd68$ has been detected
from the modulation of a number of line profile parameters,
and is in good agreement with half the magnetic rotation period $P_{\rm m}/2 =21\fd038$.
It is of great importance to apply the same procedure to other Herbig~Ae/Be stars to
determine their rotation periods and to probe the structure of their accretion regions
(see e.g.\ Table~2 of Hubrig et al.\ \cite{Hubrig2015}).

\begin{acknowledgements}
We would like to thank the ESO education and Public Outreach Department and especially
Mafalda Martins for providing Fig.~\ref{fig:sketch}.
This work was supported by the Basic Research Program of the Presidium of the
Russian Academy of Sciences P-41 and the Program of the Department of Physical
Sciences of the Russian Academy of Sciences P-17.
N.A.D. acknowledges the support of FAPERJ, Rio de Janeiro, Brazil,
for Visiting Researcher grant E-26/200.128/2015 and the St.~Petersburg State University for research grant 6.38.18.2014.
\end{acknowledgements}

\begin{appendix}

\section{The X-shooter data set}

\begin{figure}
\centering
\includegraphics[width=0.45\textwidth]{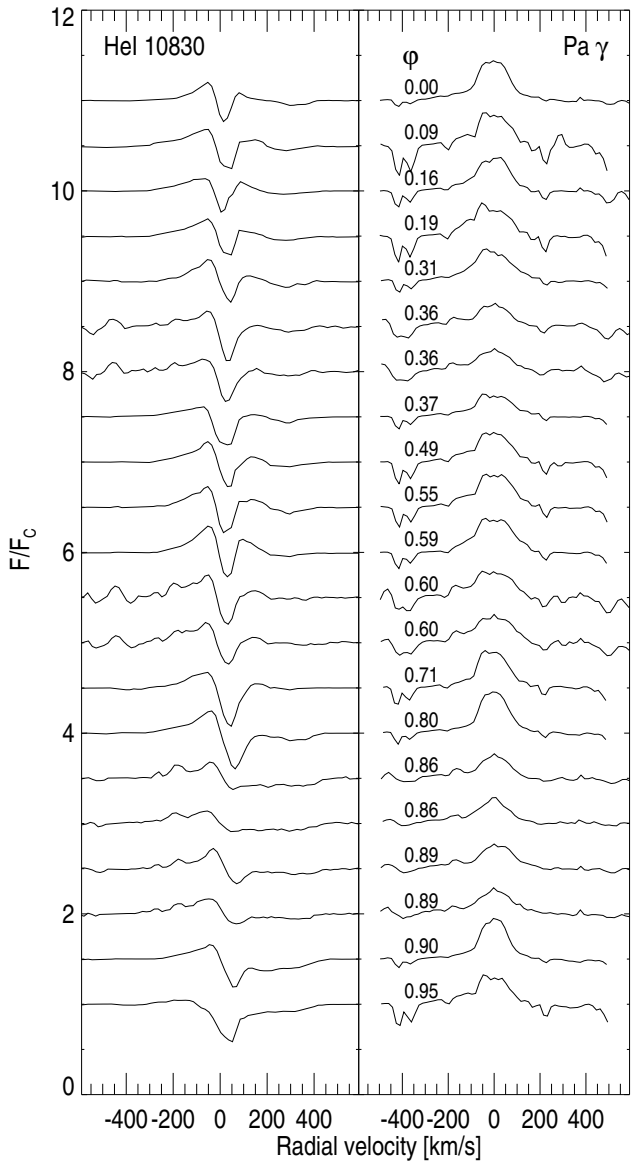}
\caption{
Line profiles of the  \ion{He}{i}~$\lambda$10,830 and Pa$\gamma$ lines in
HD\,101412 X-shooter spectra.
The rotation phases of the observations assuming the magnetic rotation
period $P_{\rm m}=42\fd076$ are presented close to the Pa$\gamma$ lines.
}
\label{fig:xshooter_set}
\end{figure}

%\clearpage
\newpage

\section{Data sets and determined spectral parameters}

\begin{table*}   %Table A1
\begin{flushleft}
\centering
\caption{
List of the HD\,101412 spectra and parameters of the line profiles.
$\varphi$ is the phase of the magnetic rotation period following Hubrig et al.\ (\cite{Hubrig2011}).
EW is the equivalent width (determined as $F_{\rm line}/F_{\rm cont}-1$),
$v_{\rm r1}$ and $v_{\rm r2}$ are
the velocities of the blue and red edges of the broad absorption component of the
\ion{He}{i}~$\lambda$10,830 line profile component.
$v_{\rm red}$ is the velocity of
the red edge of the Pa$\gamma$ emission  and $v_{\rm a} $ is the velocity of the
central \ion{He}{i}~$\lambda$10,830 absorption.
The errors of the measurements are given in brackets.
}
\label{tab:tableA1}
\begin{tabular}{lccccccccc}
 \hline
\noalign{\smallskip}
\multicolumn{1}{c}{MJD} &
\multicolumn{1}{c}{Instrument} &
\multicolumn{1}{c}{S/N} &
\multicolumn{1}{c}{$\varphi$} &
\multicolumn{4}{|c|}{\ion{He}{i}~$\lambda$10,830} &
\multicolumn{2}{c}{Pa$\gamma$}  \\
& & & &
\multicolumn{1}{|c}{EW} &
\multicolumn{1}{c}{$v_{\rm r1}$} &
\multicolumn{1}{c}{$v_{\rm r2}$} &
\multicolumn{1}{c|}{$v_{\rm a}$} &
\multicolumn{1}{c}{EW} &
\multicolumn{1}{c}{$v_{\rm red}$}  \\
& & & &
\multicolumn{1}{|c}{[\AA{}]} &
\multicolumn{1}{c}{[km\,s$^{-1}$]} &
\multicolumn{1}{c}{[km\,s$^{-1}$]} &
\multicolumn{1}{c|}{[km\,s$^{-1}$]} &
\multicolumn{1}{c}{[\AA{}]} &
\multicolumn{1}{c}{[km\,s$^{-1}$]} \\
\hline \noalign{\smallskip}
55654.085  &   CRIRES  & 280 & 0.89 & 1.49 (0.02) & 136 (5)   & 465 (6)  & 47 (2) & 1.95 (0.02)  &  140 (9)  \\
55656.062  &   CRIRES  & 270 & 0.94 & 1.57 (0.02) & 180 (7)   & 465 (10) & 73 (1) & 1.45 (0.02)  &  145 (10) \\
55662.082  &   CRIRES  & 350 & 0.08 & 1.40 (0.02) & 236 (8)   & 430 (13) & 38 (2) & 1.28 (0.02)  &  167 (15) \\
55667.174  &   CRIRES  & 410 & 0.21 & 1.40 (0.02) & 191 (9)   & 350 (17) & 22 (2) & 2.48 (0.02)  &  233 (12) \\
55681.023  &   CRIRES  & 480 & 0.53 & 1.64 (0.02) & 220 (5)   & 385 (18) & 32 (3) & 2.01 (0.02)  &  240 (3)  \\
55683.058  &   CRIRES  & 420 & 0.58 & 1.65 (0.02) & 228 (5)   & 385 (18) & 17 (1) & 2.35 (0.02)  &  259 (10) \\
55947.273 & X-shooter & 180 & 0.86 & 1.04 (0.03) & 120 (30)  & 479 (30) & 56 (4) & 1.75 (0.02)  &  132 (10) \\
55947.276 & X-shooter & 120 & 0.86 & 0.94 (0.06) &  76 (25)  & 484 (25) & 45 (6) & 1.83 (0.05)  &  140 (15) \\
55948.351 & X-shooter & 160 & 0.89 & 1.01 (0.03) & 105 (15)  & 460 (15) & 68 (7) & 2.01 (0.02)  &  197 (10) \\
55948.354 & X-shooter & 120 & 0.89 & 1.11 (0.06) &  93 (30)  & 450 (30) & 70 (8) & 1.97 (0.05)  &  211 (15) \\
55978.200 & X-shooter & 130 & 0.60 & 1.49 (0.06) & 205 (20)  & 380 (20) & 31 (4) & 2.30 (0.05)  &  220 (18) \\
55978.202 & X-shooter & 100 & 0.60 & 1.29 (0.05) & 195 (30)  & 450 (30) & 34 (4) & 2.32 (0.04)  &  240 (16) \\
56010.262 & X-shooter & 160 & 0.36 & 0.75 (0.06) & 170 (20)  & 400 (24) & 34 (5) & 1.79 (0.05)  &  200 (12) \\
56010.265 & X-shooter & 110 & 0.36 & 0.79 (0.03) & 178 (28)  & 380 (15) & 28 (4) & 1.83 (0.02)  &  190 (15) \\
56361.178 &   CRIRES  & 190 & 0.70 & 2.09 (0.02) & 275 (15)  & 280 (15) & 29 (2) & 2.16 (0.02)  &  192 (15) \\
56367.283 &   CRIRES  & 280 & 0.84 & 1.39 (0.02) & 229 (7)   & 430 (9)  & 29 (2) & 1.83 (0.02)  &  178 (8)  \\
56372.185 &   CRIRES  & 260 & 0.96 & 1.44 (0.02) & 190 (6)   & 475 (7)  & 58 (1) & 1.51 (0.02)  &  140 (11) \\
56639.267 & X-shooter & 250 & 0.31 & 1.10 (0.02) & 197 (10)  & 445 (15) & 43 (5) & 2.73 (0.02)  &  225 (6)  \\
56649.336 & X-shooter & 110 & 0.55 & 1.20 (0.04) & 205 (15)  & 413 (22) & 24 (6) & 2.74 (0.03)  &  185 (11) \\
56666.302 & X-shooter & 240 & 0.95 & 0.23 (0.02) &  90 (14)  & 470 (14) & 52 (5) & 2.06 (0.02)  &  116 (4)  \\
56672.163 & X-shooter & 230 & 0.09 & 1.09 (0.03) & 197 (10)  & 415 (30) & 33 (5) & 2.33 (0.02)  &  170 (10) \\
56676.246 & X-shooter & 160 & 0.19 & 1.00 (0.03) & 186 (10)  & 415 (30) & 32 (7) & 2.30 (0.02)  &  170 (8)  \\
56684.194 & X-shooter & 220 & 0.37 & 0.65 (0.03) & 145 (12)  & 475 (30) & 31 (6) & 1.92 (0.02)  &  165 (17) \\
56689.183 & X-shooter & 320 & 0.49 & 1.02 (0.02) & 140 (6)   & 465 (30) & 36 (4) & 2.28 (0.02)  &  160 (12) \\
56693.316 & X-shooter & 250 & 0.59 & 0.87 (0.02) & 197 (7)   & 410 (30) & 32 (5) & 2.78 (0.02) & 190 (8)  \\
56698.103 & X-shooter & 150 & 0.71 & 1.53 (0.07) & 240 (16)  & 320 (25) & 43 (4) & 2.65 (0.05) & 225 (14) \\
56702.087 & X-shooter & 150 & 0.80 & 1.13 (0.04) & 160 (20)  & 450 (20) & 62 (7) & 2.82 (0.03) & 170 (14) \\
56706.088 & X-shooter & 160 & 0.90 & 0.69 (0.04) &  80 (25)  & 476 (25) & 61 (6) & 2.49 (0.03) & 160 (13) \\
56710.367 & X-shooter & 330 & 0.00 & 0.87 (0.02) & 219 (28)  & 440 (16) & 18 (6) & 3.07 (0.02) & 210 (13) \\
56717.047 & X-shooter & 310 & 0.16 & 0.93 (0.03) & 181 (18)  & 400 (30) & 10 (5) & 2.48 (0.02) & 190 (11) \\
\noalign{\smallskip} \hline
\end{tabular}
\end{flushleft}
\end{table*}

\end{appendix}


\begin{thebibliography}{}

\bibitem[2008]{Alecian2008}
Alecian, E., Catala, C., Wade, G.~A., et al.\ 2008,
MNRAS, 385, 391

\bibitem[2013]{Alecian2013}
Alecian, E., Wade, G.~A., Catala, C., et al.\ 2013,
MNRAS, 429, 1001

\bibitem[2015]{AlvaradoGomez2015}
Alvarado-G{\'o}mez, J.~D., Hussain, G.~A.~J., Grunhut, J., et al.\ 2015,
A\&A, 582, A38

\bibitem[2007]{Bouvier2007}
Bouvier, J., Alencar, S.~H.~P., Harries, T.~J., Johns-Krull, C.~M., \& Romanova, M.~M.\ 2007,
in Protostars and Planets V, ed.\ B.~Reipurth, D.~Jewitt, \& K.~Kaul (Tucson: Univ.\ Arizona Press), 479

\bibitem[1990]{Camenzind1990}
Camenzind, M.\ 1990,
Rev.\ in Modern Astronomy, 3, 234

\bibitem[2014]{CauleyJohnsKrull2014}
Cauley, P.~W., \& Johns-Krull, C.~M.\ 2014,
ApJ, 797, 112

\bibitem[2010]{Cowley2010}
Cowley, C.~R., Hubrig, S., Gonz\'{a}lez, J.~F., \& Savanov, I.\  2013,
A\&A, 523, 65

\bibitem[2006]{Edwards2006}
Edwards, S., Fischer, W., Hillenbrand, L., \& Kwan, J.\ 2006,
ApJ, 646, 319

\bibitem[1984]{FinkenzellerMundt1984}
Finkenzeller, U., \& Mundt, R.\ 1984,
A\&ASS, 55, 109

\bibitem[2008]{Fischer2008}
Fischer, W., Kwan, J., Edwards, S., \& Hillenbrand, L.\ 2008,
ApJ, 687, 1117

\bibitem[1998]{Gullbring1998}
Gullbring, E., Hartmann, L., Brice\~no, C., \& Calvet, N.\ 1998,
ApJ, 492, 323

\bibitem[1960]{Herbig1960}
Herbig, G.~H.\ 1960,
ApJS, 4, 33

\bibitem[2004]{Hubrig2004}
Hubrig, S., Sch\"oller, M., \& Yudin, R.~V.\ 2004,
A\&A, 428, L1

\bibitem[2009]{Hubrig2009}
Hubrig, S., Stelzer, B., Sch\"{o}ller, M., et al.\ 2009,
A\&A, 502, 283

\bibitem[2010]{Hubrig2010}
Hubrig, S., Sch{\"o}ller, M., Savanov, I., et al.\ 2010,
Astr.\ Nachr., 331, 361

\bibitem[2011]{Hubrig2011}
Hubrig, S., Mikul{\'a}{\v s}ek, Z., Gonz\'{a}lez. J.~F., et al.\ 2011,
A\&A, 525, L4

\bibitem[2013]{Hubrig2013}
Hubrig, S., Sch\"oller, M., Ilyin, I., \& Lo Curto, G.\ 2013,
AN, 334, 1093

\bibitem[2015]{Hubrig2015}
Hubrig, S., Carroll, T.~A., Sch\"oller, M., \& Ilyin, I.\  2015,
MNRAS, 449, L118

\bibitem[1961]{KendallStuart1961}
Kendall, M.~G., \& Stuart, A.\ 1961,
The Advanced Theory Of Statistics, p.~292

\bibitem[2007]{Kochukhov2007}
Kochukhov, O.\ 2007,
in Physics of Magnetic Stars, ed.\ I.~I.~Romanyuk, \& D.~O.~Kudryavtsev, 109

\bibitem[1999]{Kupka1999}
Kupka, F., Piskunov, N.~E., Ryabchikova, T.~A., Stempels, H.~C., \& Weiss, W.~W.\ 1999,
A\&A, 138, 119

\bibitem[1965]{LaeflerKinman1965}
Laefler, J., \& Kinman, T.~D.\ 1965,
ApJS, 11, 216

\bibitem[1992]{Press1992}
Press, W.~H., Teukolsky, S.~A., Vetterling, W.~T., \& Flannery, B.~P.\ 1992,
Numerical Recipes. The Art of Scientific Computing (Second Edition). Section 13.8 (C).

\bibitem[2003]{Romanova2003}
Romanova, M.~M., Ustyugova, G.~V., Koldoba, A.~V., Wick, J.~V., \& Lovelace, R.~V.~E.\ 2003,
ApJ, 595, 1009

\bibitem[2004]{Romanova2004}
Romanova, M.~M., Ustyugova, G.~V., Koldoba, A.~V., \& Lovelace, R.~V.~E.\ 2004,
ApJ, 610, 920

\bibitem[2015]{RomanovaOwocki2015}
Romanova, M.~M., \& Owocki, S.~P.\ 2015,
Space Sci.\ Rev., 191, 339

\bibitem[2000]{Shu2000}
Shu, F.~H., Najita, J.~R., Shang, H., \& Li, Z.-Y.\ 2000,
in Protostars and Planets IV, ed.\ V.~Mannings, A.~P.~Boss, \& S.~S.~Russell
(Tucson: Univ.\ Arizona Press), 789

\bibitem[1995]{Smith1995}
Smith, V.~V., Cunha, K., \& Lambert, D.~L.\ 1995,
AJ, 110, 2827

\bibitem[1994]{The1994}
Th\'e, P.~S., de Winter, D., \& P\'erez, M.~R.\ 1994,
A\&AS, 104, 315

\bibitem[2005]{Wade2005}
Wade, G.~A., Drouin, D., Bagnulo, S., et al.\ 2005,
A\&A, 442, L11

\bibitem[2009]{ZechmeisterKuerster2009}
Zechmeister, M., \& K{\"u}rster, M.\ 2009,
A\&A, 496, 577


\end{thebibliography}
\end{document}